\documentclass[conference]{IEEEtran}

\usepackage{cite}
\usepackage{amsmath,amssymb,amsfonts}
\usepackage{algorithmic}
\usepackage{graphicx}
\usepackage{textcomp}
\usepackage{xcolor}
\usepackage{mathtools}
\usepackage{caption}
\usepackage{subcaption}
\usepackage[linesnumbered,ruled,vlined]{algorithm2e}
\usepackage[printonlyused,nohyperlinks,nolist]{acronym}
\SetCommentSty{mycommfont}
\SetKwInput{KwInput}{Input}                
\SetKwInput{KwOutput}{Output}              
\def\BibTeX{{\rm B\kern-.05em{\sc i\kern-.025em b}\kern-.08em
		T\kern-.1667em\lower.7ex\hbox{E}\kern-.125emX}}

\newcommand{\signal}[1]{\mathbf{#1}}
\newcommand{\set}[1]{\mathcal{#1}}
\begin{document}
	
	\title{Learning to Communicate with Intent: An Introduction\\
		
	}
	
	\author{\IEEEauthorblockN{Miguel A. Gutierrez-Estevez,
			Yiqun Wu, and 
			Chan Zhou
			\IEEEauthorblockA{Huawei Technologies Duesseldorf GmbH, German Research Center, Munich Office, Germany}
			Email: m.gutierrez.estevez@huawei.com}}
	
	\maketitle
	
	\begin{abstract}
		We propose a novel framework to learn how to communicate with intent, i.e., to transmit messages over a wireless communication channel based on the end-goal of the communication. This stays in stark contrast to classical communication systems where the objective is to reproduce at the receiver side either exactly or approximately the message sent by the transmitter, regardless of the end-goal. Our procedure is general enough that can be adapted to any type of goal or task, so long as the said task is a (almost-everywhere) differentiable function over which gradients can be propagated. We focus on supervised learning and reinforcement learning (RL) tasks, and propose algorithms to learn the communication system and the task jointly in an end-to-end manner. We then delve deeper into the transmission of images and propose two systems, one for the classification of images and a second one to play an Atari game based on RL. The performance is compared with a joint source and channel coding (JSCC) communication system designed to minimize the reconstruction error of messages at the receiver side, and results show overall great improvement. Further, for the RL task, we show that while a JSCC strategy is not better than a random action selection strategy even at high SNRs, with our approach we get close to the upper bound even for low SNRs.
	\end{abstract}
	
	\begin{IEEEkeywords}
		Semantic/Goal-Oriented Communications, Interlligent Air Interface, Integrated Communication and Computation
	\end{IEEEkeywords}
	
	\section{Introduction}
	\label{sec:intro}
	
	The seminal work by Shannon in 1948 \cite{shannon1948mathematical} quantifying the maximum data rate that a noisy communication channel can support set the foundation of modern communication systems. In Shannon’s work, all messages were treated equally regardless of semantic meaning or final goal of the communication. This assumption, together with the guarantee that separation of source and channel coding performs equally as a \ac{JSCC} strategy in the infinite blocklength regime, motivated the split of the communication and the application into two independent systems. This principle has defined the way all communications systems are designed until today.
	
	But the advent of machines with intelligence and communication capabilities is starting to reveal the limitations of the classic theory. Communication systems have been designed up until now with the objective of reconstructing the messages at the receiver side with the highest fidelity possible, because a human has always been assumed to be the ultimate consumer of information. With machines however, fidelity of reconstructed messages is not necessarily the most relevant criterion to guarantee optimal operation. Consider for example a robotic application where a robot collects sensing information and sends it to a central server for processing, after which the output is sent back to the robot. Depending on the task, e.g. object detection, there are clearly some segments of the sensor information more relevant for the optimal performance of the task than others, such as areas on the video feed of the robot where relevant objects lie. An optimal communication system for such a task should be designed with the awareness of this contextual relevance.
	
	Shannon and Weaver famously identified three levels of problems within the broad subject of communication \cite{shannon1959mathematical}: The technical level, the semantic level, and the effectiveness level. In this work, we focus on the effectiveness level of communications by proposing learning mechanisms to design \ac{GOCom} systems. Inspired by the recent success of deep learning techniques for the design of \ac{JSCC} systems for the transmission of images \cite{bourtsoulatze2019deep,xu2021wireless}, text \cite{farsad2018deep} or even videos \cite{tung2022deepwive}, we design a \ac{GOCom} system where an encoder and a decoder separated by a wireless channel are jointly learned in order to generate a task output based on an input signal (see Fig. \ref{fig:gocomms_system_model} for reference). Our framework is general enough to handle any kind of learning task and communication channel, as long as the task and the channel are (almost-everywhere) differentiable functions. We introduce algorithms for both supervised learning and \ac{RL}, and we present the case study of \ac{GOCom} for image transmission. Within the case study, we design two communications systems, the first one focused on an image classification task, while the second one is focused on a \ac{RL} task. We show with simulation results that the intuition behind designing specialized communications systems for a particular task indeed holds and \ac{GOCom} increases performance when compared with \ac{JSCC}, especially in bad channel conditions. Moreover, we show that for the application of playing the Atari game BreakOut with \ac{RL}, the agent is extremely sensitive to the distortion of reconstructed signals and fails drastically with a \ac{JSCC} strategy, while by using our \ac{GOCom} approach, the communication system is able to focus on the relevant parts of the transmitted information.
	
	\textit{Prior Art:} \ac{GOCom} is starting to gain a great deal of attention. A research direction \cite{strinati20216g,lan2021semantic} is focused on the architectural challenges, layer structure and new applications towards a 6G system supporting these new types of communication. Another research direction is studying particular applications that benefit from \ac{GOCom}, such as \ac{MARL} environments \cite{tung2021effective}, remote image retrieval \cite{jankowski2020wireless}, or computational offloading of robotics tasks in a cloud server \cite{nakanoya2021co}. To the best of our knowledge, this is the first work that proposes a unified framework in \ac{GOCom} for any type of input signal and for both supervised learning and \ac{RL} tasks.
	
	
	\section{The \ac{GOCom} System Model}
	\label{sec:goc}
	We aim to design a communications system specific for a particular goal or task, a problem setting usually known as \ac{GOCom}. In this section, we present the system model of such a communication system together with that of a \ac{JSCC} system, another modality of communication that has regained in recent years a great deal of attention thanks to deep learning.
	
\begin{figure}
		\centering
		
		\begin{subfigure}[b]{.47\textwidth}
			\centering
			\includegraphics[width=1\textwidth]{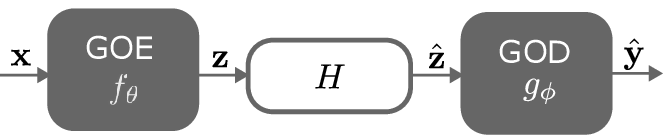}
			\caption{\ac{GOCom} system model.}
			\label{fig:gocomms_system_model}
		\end{subfigure}
		\hfill
		\begin{subfigure}[b]{.47\textwidth}
			\centering
			\includegraphics[width=1\textwidth]{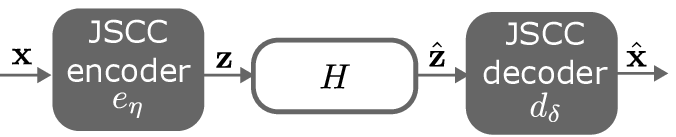}
			\caption{\ac{JSCC} system model.}
			\label{fig:jscc_system_model}
		\end{subfigure}
		\hfill
		\begin{subfigure}[b]{.47\textwidth}
			\centering
			\includegraphics[width=1\textwidth]{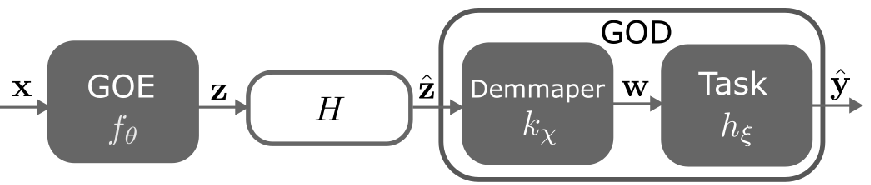}
			\caption{\ac{GOCom} system model with $\phi=[\chi,\xi]$.}
			\label{fig:gocomm_demapper_system_model}
		\end{subfigure}
		\label{fig:system_model}
		\caption{System model.}
	\end{figure}

	Let $h$ be any differentiable task mapping an input signal $\signal{x}\in\set{X}$ into an output $\signal{y}\in\set{Y}$, i.e. $h:\set{X}\rightarrow\set{Y}$. Such a task can be e.g. image classification, action selection, or path planning, just to name a few. Let $H$ be the transfer function of a differentiable channel model such as AWGN or Rayleigh (block) fading channel, which we loosely call the wireless channel hereafter. The goal in \ac{GOCom} is to transmit the signal $\signal{x}\in\set{X}$ over the wireless channel $H$ with the intent to perform task $h$ at the receiver side. To this end, the transmitter encodes signal $\signal{x}$ into an encoded representation $\signal{z}\in\set{Z}$, which is transmitted through $H$. The encoding procedure is done by what we call a \ac{GOE} represented by $f:\set{X}\rightarrow\set{Z}$. At the other side, the receiver receives a corrupted version $\hat{\signal{z}}$ of the transmitted signal, which is input to a \ac{GOD} $g:\set{Z}\rightarrow\set{Y}$ to obtain the task output $\hat{\signal{y}}$. The key difference between \ac{GOCom} and classical communications systems is that, unlike in classical communications, the decoder $g$ in \ac{GOCom} is not designed to invert the encoding function $f$ but to directly transform the received signal $\hat{\signal{z}}$ into the output $\signal{y}$ of the task $h$. Figure \ref{fig:gocomms_system_model} shows the system model for \ac{GOCom}. The \ac{GOCom} framework thus allows us to learn specialized communication systems for each particular type of input and task.
	
	We propose to design \ac{GOCom} systems by learning mechanisms such as \acp{NN}, which implies that i) all functions, i.e. \ac{GOE} $f$, \ac{GOD} $g$ and channel $H$ need to be almost-everywhere differentiable, and ii) the learnable functions ($f$ and $g$) are parametrized by a set of parameters that can be adapted during learning. To account for the latter condition, and with a little abuse of notation, we define parameter vectors $\theta$ and $\phi$ that contain all parameters (e.g. weights of a \ac{NN}) of the \ac{GOE} and \ac{GOD}, respectively; and rewrite these functions as $f_\theta$ and $g_\phi$, respectively.
	
	Figure \ref{fig:jscc_system_model} shows the system model of a \ac{JSCC} communication system. Similarly to \ac{GOCom}, in \ac{JSCC} the encoder $e$ maps the input signal $\signal{x}$ into an encoded representation $\signal{z}$, which implicitly performs the codification of source and channel in one single step, i.e. $e:\set{X}\rightarrow\set{Z}$. The difference comes at the receiver side where the decoder $g$ inverts the encoding operation with the objective to reconstruct the original transmit signal $\signal{x}$, i.e. $d:\set{Z}\rightarrow\set{X}$. As with \ac{GOCom}, we assume that both encoder and decoder are parametric functions that can be learned, and thus we collect their parameters in vectors $\eta$ and $\delta$, respectively. 
	In \ac{JSCC}, as well as in conventional communications systems, the communication can be seen as a goal-agnostic procedure where the objective is to reconstruct at the receiver the input signal $\signal{x}$ with the highest fidelity as possible, regardless of the final intent of the communication.
	
	An important aspect of designing \ac{GOCom} systems is the relationship between $h_\xi$, and $f_\theta$ and $g_\phi$. On could consider a communication system in which the set of task parameters $\xi$ might be split into two subsets $\xi_1$ and $\xi_2$ such that $\xi=[\xi_1,\xi_2]$, and assign $\xi_1$ as the parameters of the transmitter, while $\xi_2$ are the parameters of the receiver, i.e., $\theta=\xi_1$, and $\phi=\xi_2$. This would be analogous to conventional split learning with analog transmission where a large learning model is split in two parts, and each one is computed at different devices. While an interesting research direction, there is an important disadvantage to this scheme: Large models tend to be asymmetric in the way complexity is distributed across layers, and the layer sizes tend to be large at the beginning, while reducing in deeper layers. This has the implication that the communication requirements for the same task using different splitting points can differ greatly, and there is little control over the communication rate of such system. Instead, we propose another way of designing \ac{GOCom} systems in which, upon selecting a bandwidth compression rate $r$, i.e. the ratio between the number of transmit symbols and the size of the input signal, the \ac{GOE} function $f_\theta$ is designed to encode the input signal $\signal{x}$ into the transmit signal $\signal{z}$ so the rate $r$ is satisfied. At the receiver side, the \ac{GOD} function $g_\phi$ is split into two functions; a first function $k_\chi$ de-maps the received signal $\hat{\signal{z}}$ into another space $\set{W}$, and the output $\signal{w}\in\set{W}$ is used as input to a second function $l_\mu:\set{W}\rightarrow\set{Y}$ to finally produce $\hat{\signal{y}}$. Note that if $\set{W}$ has the same cardinality as $\set{X}$, and the same task function $h_\xi$ is considered as $l_\mu$, i.e., $l_\mu:=h_\xi$, we can write $g_\phi$ as $g_\phi:=h_\xi\circ k_\chi$, where $\circ$ is the composition operator and $\phi=[\chi,\xi]$. Figure \ref{fig:gocomm_demapper_system_model} shows the resulting architecture of our approach. In this case, we can reuse a pre-trained model for a particular task $h$ as initialization for $\chi$ before learning the \ac{GOCom} system end-to-end. 
	
	
	In this study, we focus on two different channel models, i.e. AWGN and slow fading channel. The transfer function of the AWGN channel is given by $H_a(\signal{z})=\signal{z}+\signal{n}$, where $\signal{n}\in\mathbb{C}^s$ represents \ac{i.i.d.} samples from a circularly symmetric complex Gaussian distribution, i.e., $\signal{n}\sim\set{C}\set{N}(0,\sigma^2)$, where $\sigma^2$ is the average noise power. For slow fading, we use a Rayleigh distribution, given by the transfer function $H_r(\signal{z})=c\signal{z}$, where $c\sim\set{C}\set{N}(0,H_r)$ is a complex random variable. The transfer function of slow fading with AWGN noise is thus given by $H_s(\signal{z})=c\signal{z}+\signal{n}$. We assume perfect channel knowledge at the receiver side.
	
	
	\section{Types of Tasks and their Implementation in \ac{GOCom}}
	\label{sec:tasks}
	We introduce in this section the types of tasks that our framework is able to deal with, and how to implement them in a \ac{GOCom} system.
	
	Let $c_{\theta,\chi,\xi}:\set{X}\rightarrow\set{Y}$ be the function representing the communication system in Fig. \ref{fig:gocomm_demapper_system_model} with input $\signal{x}$ and output $\hat{\signal{y}}$. Mathematically speaking, $c_{\theta,\chi,\xi}$ is the composition of $f_\theta$, $H$, $k_\chi$ and $h_\xi$, i.e. $c_{\theta,\chi,\xi}:=h_\xi \circ k_\chi \circ H\circ f_\theta$. Because \ac{GOCom} systems can be trained for a variety of tasks, from e.g. classification to \ac{RL}, we introduce a generic notion of the learning problem $\set{P}$ for a particular task as follows: A problem 
	$\set{P}=\{c_{\theta,\chi,\xi},\set{O},p,q_{\textup{init}},q(\signal{x}_{t+1}|\signal{x}_{t},\signal{y}_{t}),T\}$ 
	consists of a \ac{GOCom} function $c_{\theta,\chi,\xi}$, an objective function $\set{O}$, a channel distribution $p$, an initial distribution over all observations $q_{\textup{init}}$, a transition distribution $q(\signal{x}_{t+1}|\signal{x}_{t},\signal{y}_{t})$ with $t=1,...,T$ being the step index of an episode, and an episode length $T$. In supervised learning problems, the length is $T=1$. The objective function $\set{O}$ provides problem-specific evaluation of the performance, which might be a misclassification loss in a classification problem, a distance loss in a regression problem, or a reward of a Markov decision process in the case of \ac{RL}. Regardless of the task, our system model is general enough to deal with any kind of learning problem and channel model. We only assume that the objective function is smooth enough in the set of parameters $\theta$, $\chi$ and $\xi$ while the channel model is differentiable, so we can use gradient-based methods.
	
	For the objective $\set{O}$ of the learning problem $\set{P}$, we propose the linear combination of two terms: The first term is related to the minimization (maximization) of the task $h_\xi$ at the receiver side, whereas the second term is a regularization term bringing the output of the demapping function $k_\chi$ closer to the transmit signal $\signal{x}$. Mathematically speaking, the objective is defined as:
	\begin{equation}
		\label{eq:objective}
		\set{O}(\signal{x},\signal{w},\signal{y})=(1-\alpha)\set{L}_{\textup{task}}(\signal{x},\signal{y})+\alpha\set{L}_{\textup{comm}}(\signal{x},\signal{w}),
	\end{equation}
	where $\set{L}_{\textup{task}}$ is the task objective loss function, $\set{L}_{\textup{comm}}$ is a distance metric, and $\alpha\in[0,1]$ is a scalar balancing the weight of the task and the reconstruction functions in the objective. 
	
	Depending on the task, the objective can be the minimization of a loss function (e.g. image classification) or the maximization of a utility function (e.g. \ac{RL}). We take the convention of defining the task problem as a loss function, so in the case of a maximization problem as in \ac{RL}, we simply negate the task objective in the optimization problem. The reason for including the communication loss $\set{L}_{\textup{comm}}$ into the overall objective is a two-fold one: 
	
	First, the set of task parameters $\xi$ can be much larger than the set of parameters related to the communication functions ($\theta$ and $\chi$), in particular for complex tasks that require large \acp{NN} with millions or even billions of parameters. In such cases, the \ac{GOCom} model can very quickly overfit to the training set and channel, degrading performance greatly. Adding the communication loss $\set{L}_{\textup{comm}}$ as regularization term helps combat the overfitting in this kind of scenarios. 
	
	Second, by adding the term $\set{L}_{\textup{comm}}$, we can help with explainability issues that a system that jointly learns to communicate and to perform a task end-to-end might present. This is achieved because, by adding $\set{L}_{\textup{comm}}$ to the problem, we encourage the system to find representations of $\signal{w}$ close enough to $\signal{x}$, thus making the intermediate representations more explainable to the human eye.
	
	
	As mentioned earlier, although \ac{GOCom} systems can be used in principle for any differentiable task $h$, we focus here in the two most predominant learning techniques: supervise- and reinforcement learning. In the following, we formally define the optimization problem and present an algorithm to train a \ac{GOCom} system for both learning techniques. We finalize the section describing how to initialize the system parameters.
	
	\subsection{Supervised Learning}
	\label{subsec:supervised_learning}
	
	In supervised learning, one has a labeled dataset $\set{D}=\{\signal{x}_i,\signal{y}_i\}_{i=1}^n$ containing $n$ pairs of input $\signal{x}$-label $\signal{y}$ and the goal is to learn a function able to map the inputs to the outputs, while generalizing to unseen pairs (for supervised learning, we omit index $t$ since horizon $T=1$. The sub-index here refers to sample index). Problems can be of regression nature, where the goal is to predict the outputs of a continuous-valued function, or of classification nature, where the goal is to determine the class, i.e., an integer value, to which the inputs belong to. 
	
	More formally, inputs $(\signal{x}_i)_{i=1}^n$ are generated according to the distribution $q_{\textup{init}}$, and the task loss is represented by the expectation of the regularized error between the model output for $\signal{x}_i$ and the corresponding target value $\signal{y}_i$ for that observation, channel realization and task. The problem can be defined as follows:
	\begin{equation}
		\label{eq:problem_supervised_learning}
		\underset{\theta,\xi,\chi}{\textup{min}}~\mathbb{E}_{\signal{x},\hat{\signal{z}}\sim q,p(\cdot|\signal{z})}\left[(1-\alpha)\set{L}_{\textup{task}}(\signal{x},\signal{y})+\alpha\set{L}_{\textup{comm}}(\signal{x},\signal{w})\right].
	\end{equation}
	Since the dataset contains a finite number of samples, Problem \eqref{eq:problem_supervised_learning} is typically approximated by the empirical cost, yielding 
	\begin{equation}
		\label{eq:problem_supervised_learning_approx}
		\underset{\theta,\xi,\chi}{\textup{min}}~\frac{1}{n}\sum_{\signal{x}_i,\signal{y}_i\sim \set{D}}\!\left[(1-\alpha)\set{L}_{\textup{task}}(\signal{x}_i,\signal{y}_i)+\alpha\set{L}_{\textup{comm}}(\signal{x}_i,\signal{w}_i)\right],
	\end{equation}
	where $(\forall i\in \{1,...,n\}),~\signal{w}_i=k_\chi(\hat{\signal{z}}_i)$ is the intermediate output of the communication module, $\signal{z}_i=f_\theta(\signal{x}_i)$ is the transmit signal, and $\hat{\signal{z}}_i\sim p(\hat{\signal{z}}|\signal{z}_i)$ is a random variable drawn from the channel distribution given the transmit signal $\signal{z}_i$. Given a dataset $\set{D}$ and a channel distribution $p$, we can jointly learn $\theta,\xi,$ and $\chi$, as detailed in Algorithm \ref{alg:supervised_learning}. The $\textup{OPT}$ function in line $9$ of Algorithm \ref{alg:supervised_learning} refers to any optimizer that can be used to update the parameters, such as \ac{SGD} or the Adam optimizer.
	\begin{algorithm}
		
		\DontPrintSemicolon
		
		\KwInput{regularization parameter $\alpha$, learning rate $\lambda$, dataset $\set{D}$}
		
		\textbf{Init} $\theta,$ $\chi$ and $\xi$\\ 
		\While{not done}
		{
			Sample mini-batch $\set{D}_m$ with $m$ datapoints from $\set{D}$\\
			\tcp*{Forward pass}
			Encode: $(\forall i\in\set{D}_m),~\signal{z}_i\gets f_\theta(\signal{x}_i)$\\
			Transmit: $(\forall i\in\set{D}_m),~\hat{\signal{z}_i}\sim p(\signal{z}|\signal{x}_i)$\\
			Demap: $(\forall i\in\set{D}_m),~\signal{w}_i\gets k_\chi(\hat{\signal{z}_i}$)\\
			Task: $(\forall i\in\set{D}_m),~\hat{\signal{y}_i}\gets h_\xi(\signal{w}_i)$\\
			\tcp*{Backward pass}
			Compute gradients: $(\forall i\in\set{D}_m),~\nabla_{\theta,\chi,\xi}\set{O}(\signal{x}_i,\signal{y}_i,\signal{w}_i)$ \\
			Update parameters: $(\forall i\in\set{D}_m),~\theta,\chi,\xi\gets \textup{OPT}(\theta,\chi,\xi,\nabla_{\theta,\chi,\xi}\set{O}(\signal{x}_i,\signal{y}_i,\signal{w}_i),\lambda)$\\
		}
		
		\KwOutput{$\theta$, $\chi$, $\xi$}	
		\caption{\ac{GOCom} for supervised learning}
		\label{alg:supervised_learning}
	\end{algorithm}
	
	
	\subsection{Reinforcement Learning}
	\label{subsec:rl}
	
	In a \ac{RL} environment, the learning problem $\set{P}$ contains an initial state distribution $q_{\textup{init}}(\signal{x}_1)$  and a transition distribution $q(\signal{x}_{t+1}|\signal{x}_t,\signal{y}_t)$ mapping the probability of transitioning to state $\signal{x}_{t+1}$ given current state $\signal{x}_{t}$ and action $\signal{y}_{t}$. Together with the distribution function, the problem defines an objective $\set{O}$ and a reward function $R_t$. The problem is therefore a \ac{MDP} with horizon $T$, where the agent is allowed to query a certain number of sample trajectories for learning. The model learned $c_{\theta,\chi,\xi}$ can be seen as a policy that maps from states $\signal{x}_t$ to a distribution over actions $\signal{y}_t$ at each timestep $t=1,...,T$. Following \eqref{eq:objective}, we can now define the reward at time instant $t$ of a \ac{GOCom} system for a RL task as:
	\begin{equation}
		\label{eq:objective_rl_t}
		\hat{R}(\signal{x}_t,\signal{y}_t,\signal{w}_t)=-(1-\alpha)R(\signal{x}_t,\signal{y}_t)+\alpha\set{L}_{\textup{comm}}(\signal{x}_t,\signal{w}_t),
	\end{equation}
	and the loss for problem $\set{P}$ and policy $c_{\theta,\chi,\xi}$ is given by:
	\begin{equation}
		\label{eq:eq_rl_problem}
		\set{O}(c_{\theta,\chi,\xi})=\mathbb{E}_{\signal{x}_t,\signal{y}_t,\hat{\signal{z}}_t\sim q,c,p(\cdot|\signal{z})}\left[\sum_{t=0}^{T-1}\gamma^{t}\hat{R}(\signal{x}_t,\signal{y}_t,\signal{w}_t)\right]\!,
	\end{equation}
	where $\gamma$ is the discount. In case that the action space $\set{Y}$ is discrete, we can resort to standard \ac{RL} algorithms for discrete action spaces such as \ac{DQN} \cite{mnih2015human}, while in the case of continuous action space with deterministic policy we can use continuous policy gradient methods such as \ac{DDPG} \cite{lillicrap2015continuous}. 
	
	In the case of \ac{RL} algorithms which exploit a replay buffer as in the aforementioned works or in most of modern \ac{RL} algorithms, we need to modify the replay buffer for \ac{GOCom} systems. More precisely, a replay buffer $\set{B}$ stores tuples $(\signal{x}_t,\signal{y}_t,R_t,\signal{x}_{t+1})$ at timestep $t$ so they can be later on drawn randomly to guarantee that samples are \ac{i.i.d.}, an important condition to reduce probability of divergent behavior of \ac{RL} algorithms. We need to modify the buffer replay in \ac{GOCom} systems by also adding $\signal{w}_t$ to the replay buffer, and substituting the reward $R_t$ at time instant $t$ for the modified reward $\hat{R}_t:=\hat{R}(\signal{x}_t,\signal{y}_t,\signal{w}_t)$. The replay buffer then stores tuples of the form $(\signal{x}_t,\signal{y}_t,\signal{w}_t,\hat{R}_t,\signal{x}_{t+1})$. Other strategies for modern \ac{RL} algorithms such as $\epsilon$-greedy exploration or the inclusion of target and critic networks need no modification in \ac{GOCom}. Algorithm \ref{alg:rl} shows the learning procedure for \ac{RL} tasks.
	\begin{algorithm}
		
		\DontPrintSemicolon
		
		\KwInput{regularization parameter $\alpha$, learning rate $\lambda$, dataset $\set{D}$}
		
		\textbf{Init} $\theta,$ $\chi$ and $\xi$. Optionally target and critic networks \\ 
		\While{not done}
		{
			Sample $m$ trajectories $\set{D}_m=\{(\signal{x}_1,\signal{y}_1,...,\signal{x}_T)_i\}_{i=1}^m$ from $\set{D}$ using $c_{\theta,\chi,\xi}$\\
			\tcp*{Forward pass}
			Encode: $(\forall i\in\set{D}_m,\forall t=1,...,T),~\signal{z}_{i,t}\gets f_\theta(\signal{x}_{i,t})$\\
			Transmit: $(\forall i\in\set{D}_m,\forall t=1,...,T),~\hat{\signal{z}}_{i,t}\sim p(\signal{z}_{i,t}|\signal{x}_{i,t})$\\
			Demap: $(\forall i\in\set{D}_m,\forall t=1,...,T),~\signal{w}_{i,t}\gets k_\chi(\hat{\signal{z}}_{i,t}$)\\
			Action: $(\forall i\in\set{D}_m,\forall t=1,...,T),~\hat{\signal{y}}_{i,t}\gets h_\xi(\signal{w}_{i,t})$\\
			(optional) Update and sample from replay buffer $\set{B}$\\
			\tcp*{Backward pass}
			$(\forall i\in\set{D}_m,\forall t=1,...,T),$ estimate gradients $\nabla_{\theta,\chi,\xi}\set{O}(c_{\theta,\chi,\xi})$ using selected \ac{RL} algorithm\\
			Update parameters: $(\forall i\in\set{D}_m,\forall t=1,...,T),~\theta,\chi,\xi\gets \textup{OPT}(\theta,\chi,\xi,\nabla_{\theta,\chi,\xi}\set{O}(\signal{x}_i,\signal{y}_i,\signal{w}_i),\lambda)$\\
			(optional) Update target network(s), critic network
		}
		
		\KwOutput{$\theta$, $\chi$, $\xi$}
		\caption{\ac{GOCom} for \ac{RL}}
		\label{alg:rl}
	\end{algorithm}
	
	\subsection{Parameters Initialization}
	\label{subsec:initialization_strategies}
	
	
	For the initialization of parameters in line 1 of both Algorithms \ref{alg:supervised_learning} and \ref{alg:rl}, we first train the task model $\xi$ of task $h_\xi$ without the encoding and transmission of signals $(\signal{x}_i)_{i=1}^n$ over the wireless channel. This represents the scenario where the transmitter performs the task locally, and it can be seen as the performance upper bound of a \ac{GOCom} system. The trained model is identified as $\xi_\textup{pre}$, and we use this model as starting point for learning a \ac{GOCom} system end-to-end. Optionally, one may fix $\xi_\textup{pre}$ during the training of the \ac{GOCom} system, therefore only updating $\theta$ and $\chi$ in steps 9 and 10 of Algorithms 1 and 2, respectively. This is useful e.g. when the task model is much larger than the encoder-demapper models, because in these cases the learning procedure may overfit very quickly if the task model is updated when running either Algorithm \ref{alg:supervised_learning} or \ref{alg:rl}. The set of parameters $\theta$ and $\chi$ are initialized randomly. Other initialization strategies are possible, e.g. by initializing the communication system with the models learned from \ac{JSCC}, but we did not observe major differences when evaluating different applications.
	
	
	We will compare in Section \ref{sec:case_study} \ac{GOCom} systems learned under this initialization strategy with a baseline where we use an independently trained \ac{JSCC} system and task, and combine them to see the performance of the task when the reconstructed output of the \ac{JSCC} communication system is used as input to the task.
	
	\section{Design and Evaluation for Image Transmission}
	\label{sec:case_study}
	
	In this section, we design and evaluate two \ac{GOCom} systems for transmission of images. The first system is designed to perform a classification task of images, while the second one transmits the images to perform a \ac{RL} task with discrete action space at the receiver. 
	
	\subsection{\ac{GOCom} for Image Classification}
	\label{susec:image_class}
	
	\subsubsection{System Design}
	
	The first application that we study is image classification. To this end, we use the CIFAR10 dataset, a dataset comprising of 60k images of ten different classes. The images are three-dimensional signals with size $32\times32\times3$ pixels, where the first two dimensions indicate the height and width of the images, respectively, and the third dimension identifies the RGB color scheme. The training dataset contains 50k images, while the test dataset has 10k.
	
	For the task function $h_\xi$, we use the ResNet50 NN commonly used for image classification tasks \cite{he2016deep}. 
	The ResNet50 network is pre-trained using the ImageNet dataset, a very large dataset containing around 14 million samples an 20k labels. 
	
	We use transfer learning to adapt the ResNet50 trained on ImageNet to the CIFAR10 dataset. Concretely, we remove the last layer of ResNet50, i.e., the layer in charge of generating the class of an image, and replace it with a small \ac{NN} consisting of a 2D average pooling layer, followed by a flattening layer, and three dense layers with 1024, 512 and 10 neurons each. The first two dense layers have the ReLu function as non-linear function, while the last layer has the sigmoid function in charge of generating the probabilities of each image belonging to each class. We also add a pre-processing block before the ResNet50 network consisting on the normalization of pixel values between $[0,255]$ into $[0,1]$, and applying 2D up-sampling layer with up-sampling factor of 2. This helps the network improve its performance at the expense of increasing the total number of parameters. While training the task function $h_\xi$, we fix the parameters belonging to the ResNet50 \ac{NN} and only update the added layers on top. This helps to avoid overfitting. The resulting set of parameters is referred to as $\xi_\textup{pre}$.
	
	For the encoder $f_\theta$ and demapper $k_\chi$ functions, we use the same architecture as in \cite{xu2021wireless}, where authors propose a \ac{JSCC} system based on \acp{CNN} with attention mechanism. The attention mechanism is introduced to input the current SNR value to the \ac{NN} and help it  operate under a much larger SNR range as done in previous studies. More concretely, they design the encoder with 5 \acp{FL} modules, i.e., blocks based on 2D convolutional layers, and 4 \acp{AF} modules, i.e., attention blocks that take the output of the previous \ac{FL} as input, together with the current SNR, and produce a scaled output according to the SNR. Each of the 4 \ac{AF} modules is introduced in between two \ac{FL} modules, thus creating an alternating structure. 
	
	Each \ac{FL} $l,~l=1,...,5$, at the encoder function $f_\theta$ consists of a convolutional layer with parameters $F_l\times F_l\times K_l |S_l$, where $F_l$ is the filter size of \ac{FL} $l$, $K_l$ filters, and $S_l$ is the stride parameter. The parameters for layers 1 to 5 are: $9\times9\times256|2$, $5\times5\times256|2$, $5\times5\times256|1$, $5\times5\times256|2$ and $5\times5\times16|2$, respectively. Following the convolutional layer, each \ac{FL} $l$ has a \ac{GDN} layer, followed by the PreLu activation function. The last \ac{FL} of the encoder $f_\theta$ does not include the PreLu function. Instead, after the last \ac{FL}, the generated real-valued signal is first converted into a complex signal with half the entries, and then it is normalized to power 1 to guarantee that the power constraints of the transmitter are satisfied, i.e. $\signal{z}=\signal{z}/(s\mathbb{E}[\signal{z}\signal{z}^*])$, where $s$ is the number of transmit symbols and $(\cdot)^*$ is the complex conjugate. 
	
	Each \ac{AF} module takes the input from the previous \ac{FL} and implements a 2D average pooling layer, to which output the SNR value is concatenated, followed by two dense layers. The first dense layer has 16 neurons, and the second one has 256. 
	
	Similarly to the encoder, the demapper function $k_\chi$ implements five \ac{FL} modules and 4 \ac{AF} modules, with each \ac{AF} module in between two \ac{AF} modules. The \acp{AF} are exactly the same as in $f_\theta$, while \acp{FL} 1 to 4 implement the \ac{IGDN} instead of the \ac{GDN} layer, followed by the PreLu activation function. The last \ac{FL}, i.e., \ac{FL} 5, implements a \ac{GDN} layer followed by the sigmoid function. The parameters of the convolutional layers 1 to 5 are $5\times5\times256|1$, $5\times5\times256|1$, $5\times5\times256|1$, $5\times5\times256|2$ and $9\times9\times3|2$, respectively.
	
	Since the input size of CIFAR10 is $32\times32\times3$ and the transmit signal has $8\times8\times16/2$ complex symbols, the compression ratio $r$ of the system is $r=512/3072=1/6$. For more details about the architecture of $f_\theta$ and $k_\chi$, we refer the reader to \cite{xu2021wireless}. 
	
	\subsubsection{Numerical Evaluation}
	
	We compare our proposed \ac{GOCom} system with a \ac{JSCC} system followed by the pre-trained task as baseline, i.e., the output of the \ac{JSCC} is input to the task. For both systems, we use Adam optimizer with learning rate $10^{-4}$. 
	
	After the initialization explained previously, the task $h_\xi$ is trained for $11$ epochs, after which it starts to overfit. The achieved performance is $89\%$ accuracy, and we deem it as the upper bound for the communication systems. 
	
	The \ac{JSCC} system $f_\theta \circ k_\chi$ is trained for $200$ epochs. The reconstruction performance of the \ac{JSCC} system is based on the \ac{PSNR}, given by $\textup{PSNR}=10\log_{10}(\textup{MAX}^2/\textup{MSE}),$
	where MAX is the max value of the input signal, 255 in our case, and MSE is the mean squared error. The system is trained in the range $[-2,20]$ dB, and the performance for several SNR values is given in Table \ref{table:psnr_jscc}.
	
	\begin{table}[htbp]
		\caption{\ac{PSNR} of reconstructed signal with \ac{JSCC}.}
		\begin{center}
			\begin{tabular}{|c|c|c|c|c|c|}
				\hline
				\textbf{Channel model}&\multicolumn{5}{|c|}{\textbf{SNR}} \\
				\cline{2-6} 
				& \textbf{0 dB} & \textbf{5 dB} & \textbf{10 dB} & \textbf{15 dB} & \textbf{20 dB} \\
				\hline
				\textbf{AWGN} & 23.82 & 27.44 & 30.57 & 32.63 & 33.56 \\
				\hline
				\textbf{Slow fading} & 22.46 & 25.07 & 26.79 & 27.62 & 27.92 \\
				\hline
			\end{tabular}
			\label{table:psnr_jscc}
		\end{center}
	\end{table}
	
	
	
\begin{figure}
		
		\centering
		
		\begin{subfigure}[b]{.24\textwidth}
			\centering
			\includegraphics[width=1\textwidth]{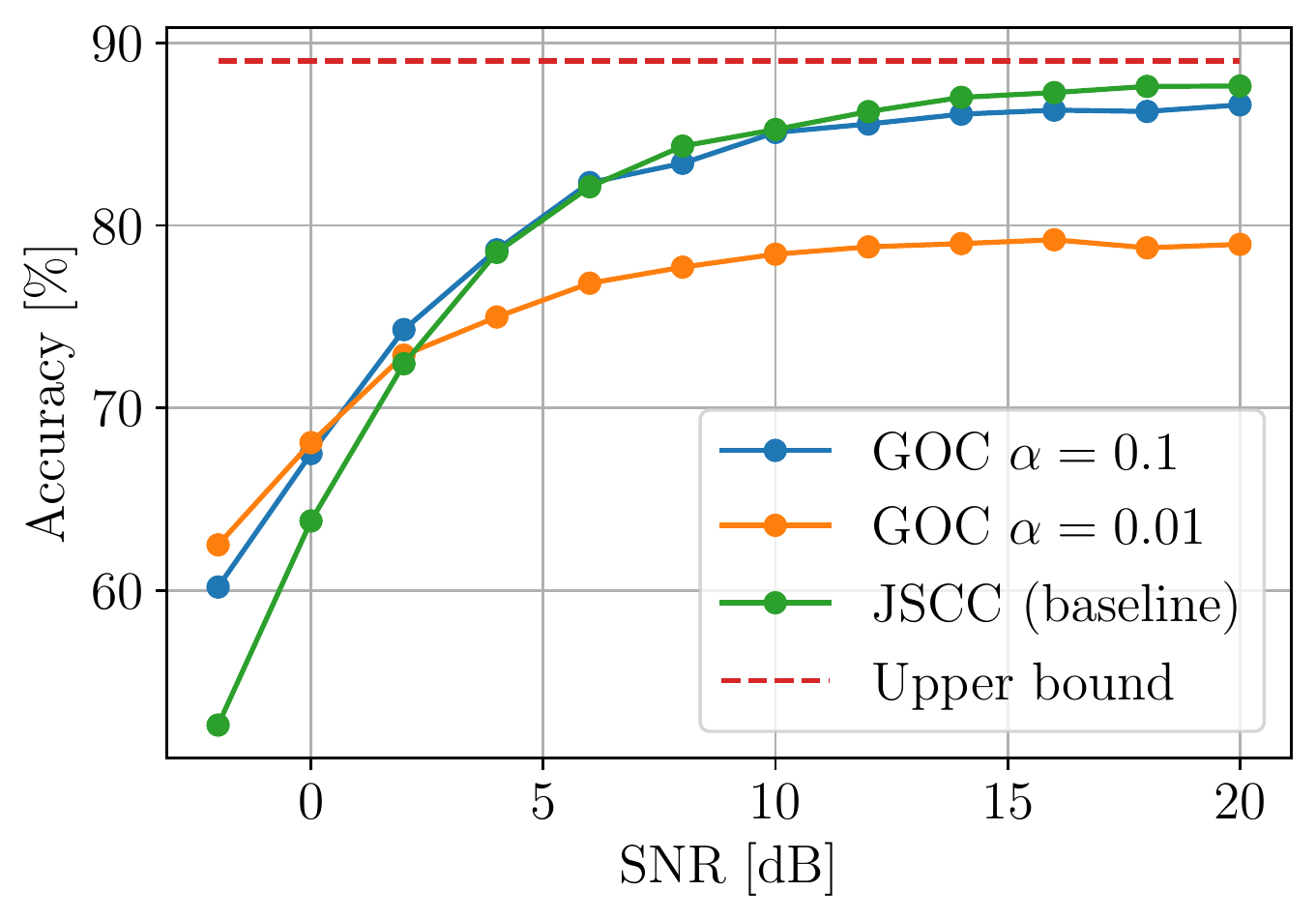}
			\caption{AWGN channel.}
			\label{fig:results_imclass_awgn}
		\end{subfigure}
		\hfill
		\begin{subfigure}[b]{.24\textwidth}
			\centering
			\includegraphics[width=1\textwidth]{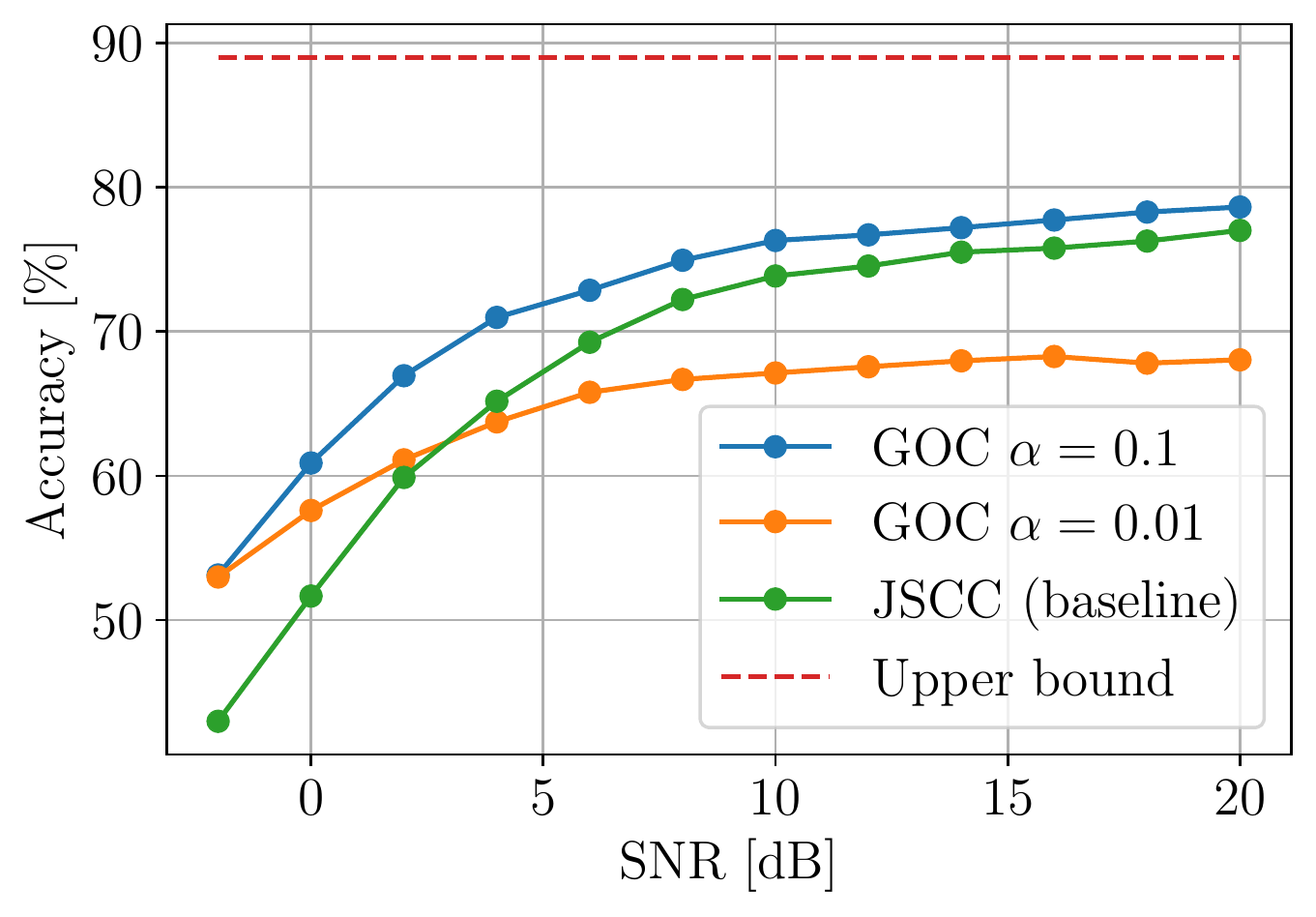}
			\caption{Slow fading channel.}
			\label{fig:results_imclass_rayleigh}
		\end{subfigure}

		\caption{Classification accuracy over SNR of CIFAR10.}
		\label{fig:results_imclass}
	\end{figure}

	All \ac{GOCom} models (GOC in Fig. \ref{fig:results_imclass}) are trained for $200$ epochs. Figure \ref{fig:results_imclass} shows the results regarding classification accuracy of both systems over the SNR for AWGN and slow fading channels. Several observations can be drawn from Fig. \ref{fig:results_imclass}. 
	\ac{GOCom} improves accuracy of the algorithm greatly in the lower SNR range compared to \ac{JSCC} (more than 10\% improvement in accuracy, or around 20\% relative improvement). For very low values of $\alpha$ (0.01), accuracy flattens at a low level for high SNRs, while in the lower SNR the accuracy is better.  
	A value of $\alpha=0.1$ seems to be a good balance, therefore obtaining good performance across the SNR range. 
	Under slow fading channel, the performance of \ac{GOCom} with $\alpha=0.1$ (Fig. \ref{fig:results_imclass_rayleigh}) is consistently better than that of \ac{JSCC}. This hints that, the more challenging the channel distribution is for the communication, the better \ac{GOCom} systems perform compared to \ac{JSCC}. The difference in performance between \ac{GOCom} with $\alpha=0.1$ and $\alpha=0.01$ also points towards the importance of adding the regularization term when learning \ac{GOCom} systems.
	
	\subsection{\ac{GOCom} for Reinforcement Learning}
	
	\subsubsection{System Design}
	
	We now focus on a \ac{RL} task that takes images as inputs and generates actions as outputs. More specifically, we use as task function $h_\xi$ the \ac{DQN} algorithm to learn how to play the Atari game BreakOut. In BreakOut, the agent controls a platform at the bottom of the screen, which can be moved left or right. The agent needs to hit a ball falling from the top, where several layers of bricks form a wall. If the ball hits any of the bricks, the score increases by one, and if the ball passes the platform at the bottom, the player loses a life. In total there are 5 lives, and the goal is to score as many points as possible before losing all lives. 
	
	Authors in \cite{mnih2015human} designed the \ac{DQN} algorithm to play Atari games, in many cases achieving super-human performance. We use the exact same algorithm and \ac{NN} as in \cite{mnih2015human}, where the input to the algorithm is a 3D signal with dimensions $84\times84\times4$, where the first two dimensions represent the height and width of the screen, while the third one is used to input four consecutive frames of the game to the algorithm. The action space is discrete with 4 possible actions, namely, move left, move right, fire a new ball, or no operation. The reward is calculated as the difference between the current score and the score in the previous frame. The \ac{NN} consists of three 2D convolutional layers with values $8\times8\times32|4$, $4\times4\times64|2$ and $3\times3\times4|1$, respectively, all layers with ReLu as non-linear function, followed by a flatten layer and two dense layers with 512 and 4 neurons respectively. For more details on the algorithm, we refer the reader to \cite{mnih2015human}.
	
	Regarding the encoder $f_\theta$, we implement a \ac{CNN} with three 2D convolutional layers with parameters $8\times8\times32|4$, $4\times4\times64|2$ and $3\times3\times192|1$, respectively. The first two layers use the PReLu activation function, while the last one implements the conversion into complex-valued output and the normalization to guarantee the power constraint. Similarly, the demapper $k_\chi$ implements three 2D transposed convolutional layers with parameters $3\times3\times192|1$, $4\times4\times64|2$ and $8\times8\times32|4$, respectively. The first two layers also implement the PReLu activation function, while the last layer uses the sigmoid function. Since the input size to the encoder is $84\times84\times4$ and the transmit signal contains $7\times7\times192/2$ complex symbols, the compression ratio $r$ of the system is $r=192\times7^2/(2\times84^2\times4)=1/6$.
	
	\subsubsection{Numerical Evaluation}
	
	We compare our proposed \ac{GOCom} system with a baseline based on a \ac{JSCC} system trained to reconstruct the 3D input signal ($84\times84\times4$ pixels) independently of the task, and then we feed the output of the \ac{JSCC} to the pre-trained task $h_{\xi_{\textup{pre}}}$, which was previously pre-trained for 10 million episodes. To guarantee a fair comparison between \ac{JSCC} and \ac{GOCom}, the encoder $e_\eta$ and decoder $d_\gamma$ components of the \ac{JSCC} system have the same architecture and number of parameters as the encoder $f_\theta$ and demapper $k_\chi$ of our \ac{GOCom} system previously introduced.
	
	The dataset to train the \ac{JSCC} system contains 12k samples randomly drawn from several BreakOut episodes, and we split the dataset in 10k samples for training and 2k samples for testing. The optimizer is Adam with learning rate $\lambda=10^{-4}$, and we train during 1028 epochs. Because there is no attention mechanism for the communication model as in the image classification task, we train two models, one at 0 dB and one at 20 dB. Table \ref{table:performance_jscc_breakout} shows the test results for several SNRs of both models under AWGN. We can see that the reconstruction accuracy is very high despite the low complexity of the \ac{NN} and the low number of samples. This is because all samples in this task are very similar to each other, so it is easy for the \ac{NN} to reconstruct them with high accuracy.
	
	\begin{table}[htbp]
		\caption{\ac{PSNR} of reconstructed signal with \ac{JSCC}.}
		\begin{center}
			\begin{tabular}{|c|c|c|c|c|c|}
				\hline
				\textbf{Train SNR}&\multicolumn{5}{|c|}{\textbf{SNR}} \\
				\cline{2-6} 
				& \textbf{0 dB} & \textbf{5 dB} & \textbf{10 dB} & \textbf{15 dB} & \textbf{20 dB} \\
				\hline
				\textbf{0 dB} & 44.87 & 45.38 & 45.50 & 45.68 & 45.74 \\
				\hline
				\textbf{20 dB} & 35.84 & 41.78 & 44.65 & 45.87 & 46.27 \\
				\hline
			\end{tabular}
			\label{table:performance_jscc_breakout}
		\end{center}
	\end{table}
	
\begin{figure}[htbp]
		\centering
		\includegraphics[width=.5\textwidth]{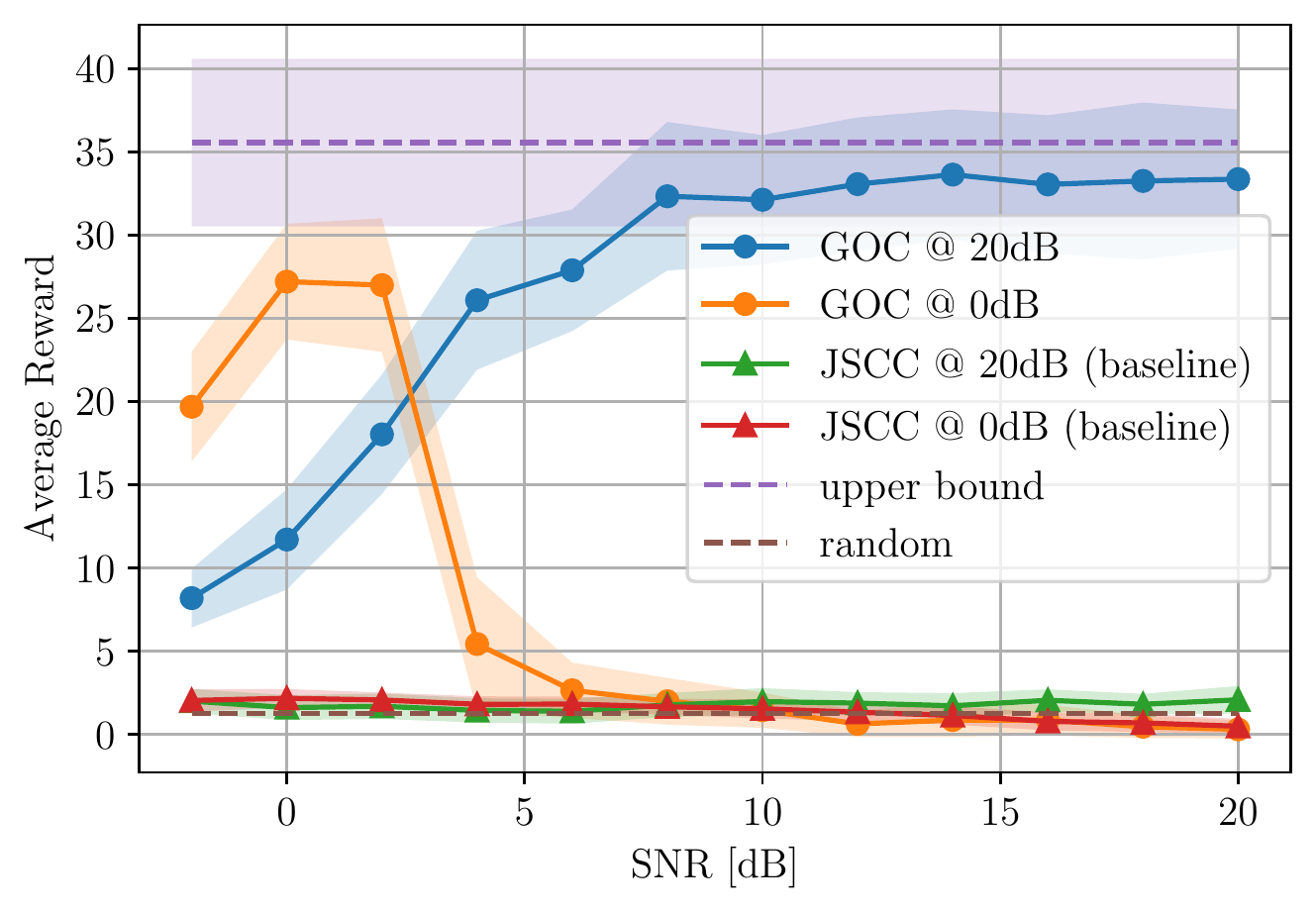}
		\caption{Rewards over SNR under AWGN for the \ac{DQN} task.}
		\label{fig:results_breakout_awgn}
	\end{figure}

	We train two \ac{GOCom} systems for 4 million episodes each under AWGN, one at 0 dB and one at 20 dB. The task is initialized  with $h_{\xi_{\textup{pre}}}$, and $\alpha$ is set to $0$ in both experiments. We evaluate the four models, i.e. \ac{JSCC} at 0 dB and 20 dB, and \ac{GOCom} at 0 dB and 20 dB, over the SNR range $[-2,20]$ dB. We run 100 episodes for each experiment at each test SNR, and plot the average reward of the 100 episodes together with its standard deviation in Fig. \ref{fig:results_breakout_awgn}. As upper bound we consider the pre-trained task $h_{\xi_{\textup{pre}}}$ with no communication system evaluated also over 100 episodes. The average reward of the upper bound is 35.57. We also plot the reward obtained by a strategy where actions are picked randomly, which achieves an average reward of $1.28$.
	
\begin{figure}
		\centering
		
		\begin{subfigure}[b]{.15\textwidth}
			\centering
			\includegraphics[width=1\textwidth]{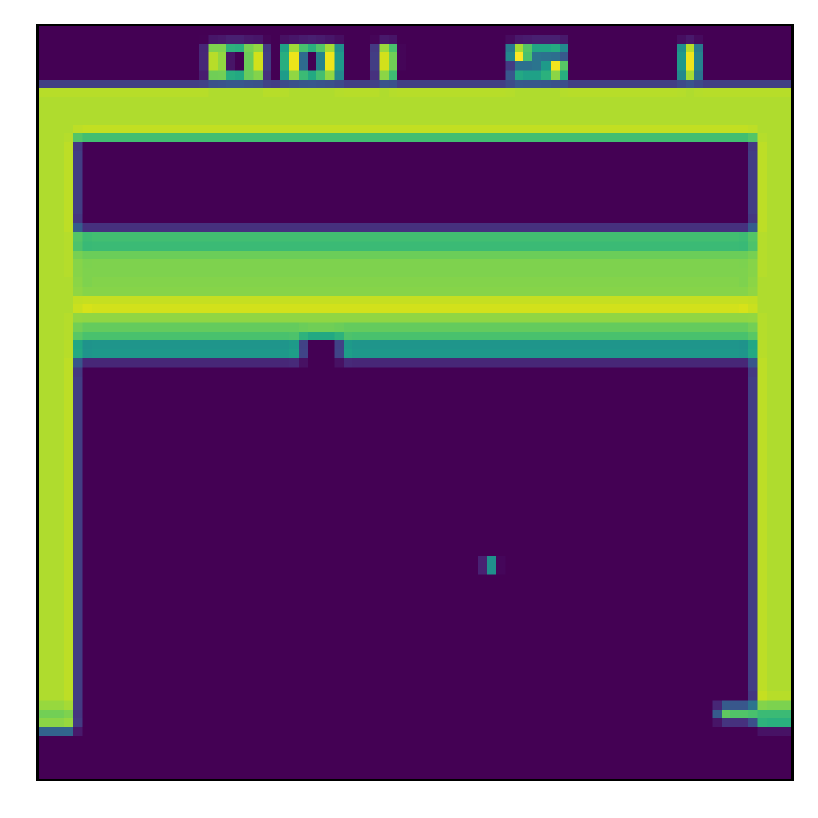}
			\caption{Input signal $\signal{x}$.}
			\label{fig:x}
		\end{subfigure}
		\hfill
		\begin{subfigure}[b]{.15\textwidth}
			\centering
			\includegraphics[width=1\textwidth]{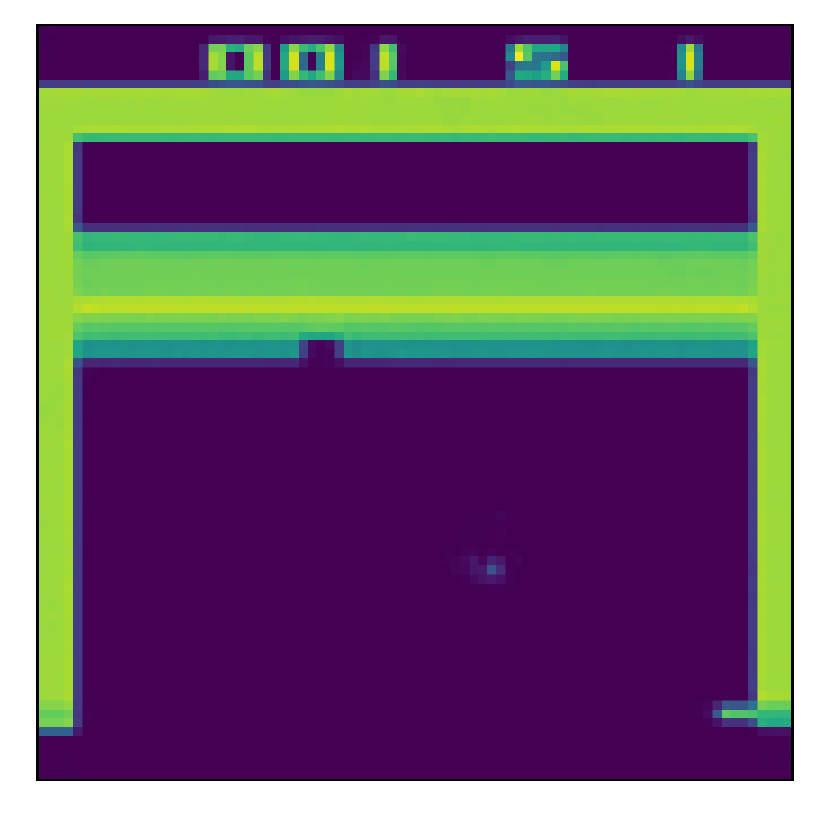}
			\caption{\ac{JSCC}.}
			\label{fig:x_hat_jscc}
		\end{subfigure}
		\hfill
		\begin{subfigure}[b]{.15\textwidth}
			\centering
			\includegraphics[width=1\textwidth]{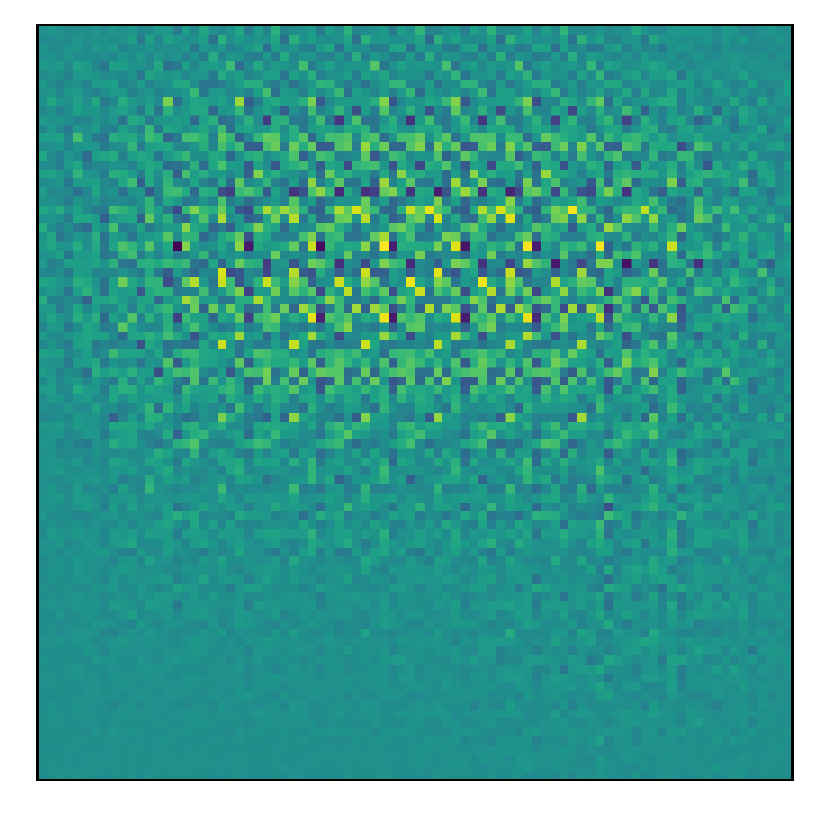}
			\caption{\ac{GOCom}.}
			\label{fig:w_goc}
		\end{subfigure}
		\label{fig:reconstructed_signals}
		\caption{Input signal (left) and output signals for \ac{JSCC} (center) and the demapper for  \ac{GOCom} (right).}
	\end{figure}

	Looking at Table \ref{table:performance_jscc_breakout} (also see Fig. \ref{fig:x_hat_jscc}), we observe that the quality of the reconstructed signal with \ac{JSCC} is extremely high, also according to human perception. However, both \ac{JSCC} systems fail drastically at the game of BreakOut, since they barely improve the random strategy (see Fig. \ref{fig:results_breakout_awgn}). Our explanation is that in \ac{JSCC}, all pixels contribute equally to the loss function, but in the game BreakOut, those pixels around the platform and the ball are more critical than the rest. If those pixels are not reconstructed with very high fidelity, the reward can be disproportionately affected. Observing Fig. \ref{fig:x_hat_jscc}, where the output of the \ac{JSCC} system trained and tested at 20 dB is depicted, one can see that, even though the quality of the reconstructed signal is very high according to the human eye, for the \ac{RL} task it is not the case, since the ball is barely appreciable. On the other hand, the \ac{GOCom} systems show a remarkable performance even at very low SNR, almost reaching the upper bound already at 8 dB for the system trained at 20 dB. However, if we look at the output $\signal{w}$ of the demapper function $k_\xi$ in Fig. \ref{fig:w_goc}, the signal has no structure for a human, which has negative impact in the explainability of the system, but the gains in terms of performance are remarkable.
	
	\section{Conclusions}
	\label{sec:conclusions}
	We have introduced a new framework to learn how to communicate with intent. The main idea relies on jointly learning how to communicate and perform a learnable task such as supervised or reinforcement learning. The intuition behind it is as follows: The relevant information in the source to perform a certain task is task-specific, and it might vary greatly between tasks. Therefore, we would like to design specialized communication systems that take this context into account when performing source and channel coding. The expected benefits are a higher efficiency in the communication and robustness against channel degradation, among others. We show in the evaluation results that the performance of such a system improves when compared with a \ac{JSCC} system, especially under challenging channel conditions. Further, for the \ac{RL} task, the performance of the \ac{GOCom} system is remarkably good, while the \ac{JSCC} system fails drastically even at high SNR values. This result points towards the need of \ac{GOCom} systems for intelligent machine interactions.

	

\begin{acronym}[SPACEEEEEE]
	\setlength{\itemsep}{-0.2em}
	\acro{GOCom}{goal-oriented communications}
	\acro{GOD}{goal-oriented decoder}
	\acro{GOE}{goal-oriented encoder}
	\acro{NN}{neural network}
	\acro{JSCC}{joint source and channel coding}
	\acro{MSE}{mean squared error}
	\acro{MAE}{mean absolute error}
	\acro{MS-SSIM}{multi-scale structural similarity index measure}
	\acro{BCE}{binary cross-entropy}
	\acro{GCE}{generalized cross-entropy}
	\acro{SGD}{stochastic gradient descent}
	\acro{RL}{reinforcement learning}
	\acro{MDP}{Markov decission process}
	\acro{DQN}{deep Q-network}
	\acro{DDPG}{deep deterministic policy gradient}
	\acro{i.i.d.}{independent and identically distributed}
	\acro{CNN}{convolutional neural network}
	\acro{AF}{attention feature}
	\acro{FL}{feature learning}
	\acro{GDN}{generalized divisive normalization}
	\acro{IGDN}{inverse GDN}
	\acro{MARL}{multi-agent RL}
	\acro{PSNR}{peak SNR}
\end{acronym}

	\bibliographystyle{IEEEtran}
	\bibliography{references}
	\addcontentsline{toc}{part}{Bibliography}
\end{document}